# vSDNEmul: A Software-Defined Network Emulator Based on Container Virtualization

Fernando N. N. Farias, Antônio de O. Junior, Leonardo B. da Costa, Billy A. Pinheiro, Antônio J. G. Abelém

Post-Graduate Program in Computer Science
Federal University of Pará – UFPA
Belém, Brazil
Email: {fernnf, ajr, lbc, billy, abelem}@ufpa.br

*Abstract*—The main issue related to Software-Defined Network emulators is how to replicate real behavior in experiments. Mininet and others SDN emulators have an architecture that limits both the scope of experiments and the fidelity of networking tests. Consequently, the serialization, contention, and load of background processes may produce delays that compromise the operation of events such as transmitting a packet or completing a computation, possibly invalidating the performance evaluation of a network emulation. To address these problems, this paper presents vSDNEmul, a network emulator based on Docker container virtualization. Different from Mininet, vSDNEmul isolates each node in a container and interconnects the nodes through virtual or tunnel links. By using containers, vSDNEmul allows autonomous and flexible creation of independent network elements, resulting in more realistic emulations. This paper reports performance evaluations comparing vSDNEmul and Mininet. The results obtained with the vSDNEmul emulator are more realistic and present higher accuracy.

## I. Introduction

In the context of next-generation networks, software-defined networking (SDN) [1] has emerged in recent years as a new concept that promises to change the limitations of current network infrastructures by breaking the vertical integration of the network by splitting the network control logic (control plane) from routers and switches that forward flow packets (data plane) [2]. This novel network model provides several benefits including more flexibility, reduction of operational costs, more efficient resource use, and easier management requirements [3]. Furthermore, as another main feature, SDN provides flexibility for extending the network with new solutions or functionalities on either the hardware or software level. However, this feature introduces a challenge for environments and facilities regarding the replication of realistic behavior of SDN production infrastructures in controlled facilities. In this case, two types of approaches are highlighted: programmable testbeds and network emulators.

Programmable testbeds offer a high level of realism by using real switches, links with high bandwidth, network elements (e.g., WiFi routers, IoT nodes, and sensors) and realistic user traffic. Facilities such as GENI and FIBRE have supported small and medium-scale experiments involving several types of resources. These resources are fully programmable, with backbone and campus connectivity at layers 3 and 2, including SDN tools. In contrast, network emulation has been widely used for developing proposals in SDN environments. By means of emulators provide ease of setup, real network behavior and low costs, these emulators have attracted attention from both the academic community and industry professionals.

Despite being a good option for prototyping SDN solutions, programmable testbeds are costly, do not represent isolated environments and present restrict scalability. In addition, the reproducibility of the experiments is limited, and the long wait times for allocating and consuming network resources are detrimental, especially in large-scale experiments with complex network topologies. In contrast, emulation allows the use of real source code in realistic networking and computing scenarios. Emulation combines the advantages of both isolated environments (e.g., simulators) and programmable testbeds for providing a solution that can be used to carry out trials, performance evaluations, protocol debugging, reproducibility of experiments, and education and research. These issues have motivated researchers to employ network emulators instead of programmable testbeds for the development of SDN applications and network solutions.

In SDN research and development, emulators such as Mininet [4], MaxiNet [5], and Mininet-WiFi [6] have been employed to exploit rich experiment scenarios through lightweight virtualization on a personal laptop. These tools became more interesting with the possibility of running the same source code in both real networks and emulations. Nevertheless, the implementation of an emulator offers no guarantee of performance fidelity due to the lack of resource isolation (e.g., Memory, CPU or I/O), which is required by hosts and switches. As a result, the overload caused by serialization, contention, and background processes can lead to delays that interfere with the operation of events such as transmitting a packet or completing a computation. Additionally, emulators based on the Mininet architecture (as mentioned previously) have other limitations such as lack of support for multiple operating systems, having hosts sharing different file systems, running new data plane protocols, and performing different SDN protocols within the same experiments. Therefore, there is a need for new emulators that offer modern components or more flexible architecture, increasing experiment fidelity and expanding the range of experiments.

To address the weaknesses identified, this paper proposes vSDNEmul, a network emulator based on container virtualization that uses the docker framework. vSDNEmul is being developed as a new emulation architecture capable of setting up network experiments with several nodes (e.g., hosts, switches or routers) with more realistic interaction and execution scenarios. The adoption of docker containers is fundamental for setting up nodes with resource isolation and full virtualization without compromising the scale of an experiment – compared with Mininet – while conserving all native lightweight virtualization. Each node in the experiment is a docker image that can be customized as a host (client or server), switch or router. The operations are completely independent and there is no sharing of a single filesystem. This paper describes performance evaluations comparing vSDNEmul and Mininet regarding CPU, memory and I/O delay.

The remainder of this paper is organized as follows. Section 2 describes the background regarding SDN and container-based emulation for network. Section 3 discusses the related work. Section 4 describes the system architecture of vSDNEmul. Section 5 discusses the performance evaluation and results obtained. Finally, Section 6 presents concluding remarks and identifies future work.

## II. BACKGROUND

### A. Software Defined Networking

SDN provides an architecture that splits a network into data and control planes, aiming to simplify the execution of network operations, reduce costs, and accelerate the evolution and deployment of new services and protocols. In the SDN model, the control plane consists of a remote element, called a controller, that has a global view of the network and centralizes the programmable operations regarding network activities, such as forwarding and dropping. The data plane is composed of network equipment that holds the minimum requirements to perform forwarding of flow data among network interfaces, whose low level actions are abstracted by control protocols such as OpenFlow [7].

The programmability and flexibility features of SDN allows for innovations in networking and has several benefits, among which we highlight:

- *physical infrastructure based on neutral/multi-vendor*: With the plane separation (control and data), network control is not embedded on network devices, reducing the dependency of the vendor-solution, which allows uniform control over multiple network devices without compatibility problem between switches and routes;
- *enables network virtualization*: SDN enables more extensive network virtualization, such as computing virtualization. Here, the abstraction applied to hardware (e.g., nodes, ports and links) allows physical infrastructure to be shared among multiple and virtual infrastructures (virtual topologies) in a controlled and isolated fashion;
- *centrally managed*: The network intelligence is logically centralized in the SDN controller (or network operating system -- NOS), which improves both flexibility and scalability by retaining a global view of the network. In addition, decoupling and centrality features permit more efficient decision making and management of the available resources;
- *programmatically configured*: SDN allows network services to configure, manage and optimize resources quickly and dynamically through self or third-party software without need for proprietary software or human intervention.

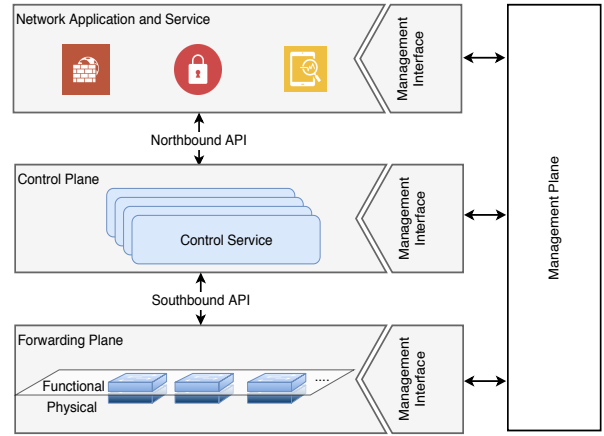

Figure 1. SDN architecture, adapted from RFC 7426 [8].

Figure 1 illustrates an SDN architecture consisting of 4 layers. The forwarding layer is composed of the physical elements such as switches and routers interconnected via wired or wireless links. Moreover, this layer manages traffic forwarding of packet flows. The control layer is responsible for managing the network intelligence and dynamics involving the collection of functions controlling one or more network devices. Furthermore, this layer has the ability to make a forward decision after processing a packet, where those decisions include forwarding, rewriting and dropping a packet. The application layer contains a set of applications and services that dictates the network behavior. The management plane is a vertical layer with functionalities including fault and monitoring management and configuration management. In addition, management-plane functionalities may include entities such as orchestrators, virtual network function managers and virtualized infrastructure managers. The northbound and southbound APIs provide interlayer abstraction. The southbound API runs protocols (e.g., OpenFlow, OVSDB and PCE) that abstract actions on physical devices [1] The northbound API provides abstraction between service control and network application through high level protocols such as restful and xmlrpc.

### B. Container-based Emulation

A network emulator represents the execution of software on a computer while configuring and running all the features that exist in a real network (i.e., switches, links, data packets, and clients/servers). The emulator runs real source code

(e.g., OS kernels, network applications, and protocols) with real network traffic. In addition, the emulator supports arbitrary topologies with a virtual "hardware" of low cost. At first glance, a network emulator with full-system virtualization is appealing. However, the use of virtual machines (VMs) is heavyweight, which can lead to more problems than solutions. The issues related to VM size and overhead may limit the scalability of nodes. The memory for each VM limits the performance of switches and hosts. Furthermore, the variability provided by a hypervisor can reduce emulation fidelity.

Container-based emulators (CBE), which use container-based virtualization techniques [9], have become popular for their efficiency and scalability advantages over hypervisor virtualization without emulating an entire computer. These emulators exploit lightweight virtualization features built for their respective operating systems, such as Linux namespaces, FreeBSD jails, and OpenSolaris zones. Because of the lack of full emulation of hardware, the software running in containers has to be compatible with the kernel and CPU architecture of the system. Containers trade the ability to run multiple OS kernels for lower overhead and better scalability than full-system virtualization. Each virtual node is simply a group of user-space processes, and the cost of adding one node is only the cost of spawning a new process.

Thus, the creation of CBE emulators has been progressing more than that of full-system virtualization, because, in addition to affording lightweight virtualization (native performance) as its main feature, the CBE technique contributes to the provisioning of other features, such as dynamic resource allocation, high density, multiple OS kernels, security isolation and filesystem isolation [10].

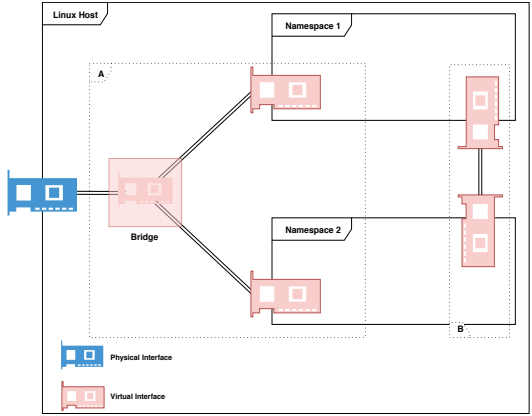

Figure 2. Use of virtual interfaces through network namespaces on Linux systems.

The Linux namespace is the most essential technology for implementation of containers for CBE emulators. The purpose of each namespace is to embed a specific system and processes isolated within the namespace. The application of namespaces is individual. Linux provides six types of namespaces; however, CBE approaches work with only process and network namespaces to implement their nodes.

Network namespaces provide isolation for the system resources associated with virtual of physical network devices.

Figure 2 illustrates the use of network namespaces on a Linux system. As illustrated in Figure 2 (box A), a namespace may have to access a physical interface or another namespace through the creation of a bridge. As depicted in Figure 2 (box B), a virtual network device provides a direct pipeline abstraction that can create tunnels between network namespaces. As another option, a physical network device can be contained in a network namespace.

One of the key properties of an SDN emulator is the use of namespaces, from which we highlight the Mininet emulator as the main existing solution. In essence, Mininet creates virtual topologies through host processes in namespaces and connects the topologies by means of virtual interfaces. Figure 3 depicts the Mininet architecture.

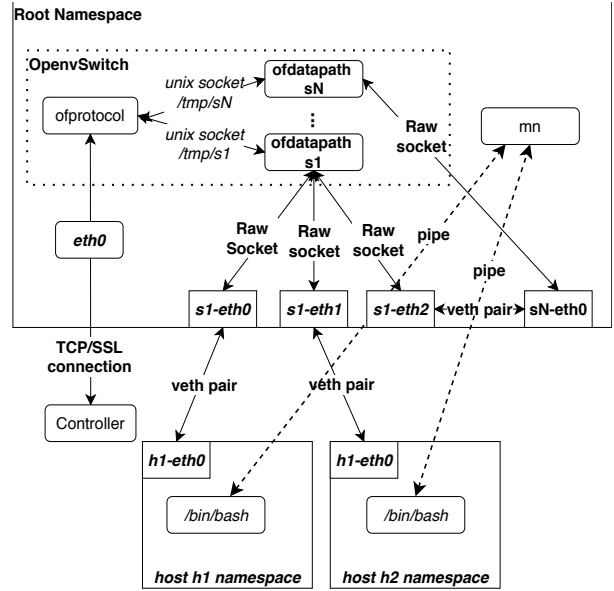

Figure 3. Mininet architecture adapted from [4].

Mininet is a popular solution for network emulation. However, the main limitation of Mininet is the low performance fidelity in experiments with high loads of network traffic. This issue is due to the namespace technology. Although Mininet has been a lightweight solution, certain issues and experiments can reduce capabilities of Mininet due to the sharing of the same resources among all the emulation elements, such as the filesystem, memory and CPU, which can influence the execution of a process inside a container, consequently affecting any experiment realized. Because of this issue, the search for new requirements in container solutions, such as Docker or LXC, facilitates experiments with more fidelity, new features, and network models.

## III. RELATED WORK

Network emulators have been applied to expand new ideas to the SDN ecosystem. The interest in emulators from both academia and industry is due to the ease of setup of these emulators, a behavior that equals a physical network, and low implementation costs. Network emulators that use container-based virtualization have become popular for efficiency and scalability advantages over full-system virtualization without emulation of an entire computer.

Lantz et al. [4] introduced Mininet, a network emulator software that enables the quick launch of a prototype virtual network with SDN switches. This lightweight virtualization uses Linux-container virtualization (namespace) features that comprise processes and network namespaces for analyzing and developing SDN solutions. The proposal presents an architecture based on namespaces where there is a root-namespace that comprises virtual switches based on Open-vSwitch [11] (to emulate SDN switches) that are linked through pair virtual Ethernet interfaces. There are also dynamic namespaces for supporting host nodes. The architecture still provides a CLI (command line interface) where a user can interact with each emulation element.

Wang et al. [12] proposed EstiNet, an OpenFlow network emulator and simulator that supports the test of functions and performance in SDN. EstiNet merges the advantages of both simulation and emulation techniques, similar to an ns-3 network simulator [13]. The simulated network of this emulator can have hosts working with a real Linux operating system. Moreover, any application of this host can run without adjustments. Therefore, this emulator can without adjustments. Therefore, this emulator can use a real SDN controller to control the simulated network devices (e.g., routers or switches).

Peuster et al. [14] created Containernet, a Mininet fork offering an expansion that supports Docker containers in place of namespaces. This proposal adds a new feature to the Mininet architecture that allows the insertion and removal of containers from the emulated network at run time. The approach of Containernet is as a cloud infrastructure, in which the computing instances can be stopped and started. Containernet provides the resource constraints, e.g., CPU time available for a single container, at run time rather than once a container has been initiated.

Hibler et al. [15] introduced a network emulator called vEmulab that virtualizes resources (e.g., hosts, routers and networks) in low-end computers, preserving high performance, significant fidelity, and near-total transparency to applications. The architecture of vEmulab uses BSD jail namespaces (analogous to Linux namespaces) to set up virtual nodes. In addition, the key design characteristics of this emulator consist of using the minimal degree of virtualization to employ the hierarchy established in real computer networks, perform optimistic automated resource allocation, and enforce observation to allocate resources in a flexible way. vEmulab is deeply automated, making this emulator easy to use even when scaling to more than a thousand virtual nodes.

In vEmulab, the authors observed the advantages of CBE over HBE systems, in which the VM size and the hypervisor overhead may limit scalability and reduce performance fidelity. In addition, these researchers have shown that CBE can be employed to reproduce an SDN environment and replicate other types of networks or solutions previously released. However, vEmulab has limitations such as lack of support to SDN and scalability constraints.

EstiNet presents a solution that integrates simulation and emulation for SDN experiments. Nevertheless, the approach has open issues. For example, the data plane features are simple and limited, which reduces performance fidelity. As a result, an EstiNet solution cannot replicate the behavior in a real SDN infrastructure. Another issue is the expansion and testing of new SDN protocols that in the proposal are conditional to the EstiNet developers.

Mininet – the most popular emulator for SDN – provides a local environment for network innovation that complements real experimentation infrastructures (e.g., FIBRE and GENI). However, there are considerable limitations that influence performance fidelity at large loads and scalability when applying complex topologies. By offering only partial virtualization, Mininet also limits its features. For example, Mininet lacks support for the simultaneous use of different operating systems, resource isolation and limitation, and high-level throughput.

With the Containernet extension, certain Mininet issues have been resolved. Containernet works with Docker containers instead of Linux namespaces, which includes features such as resource isolation and limitation. The use of prebuilt Docker images improved the admittance of modern services and applications to experiment. However, the solution was particularly restricted to host nodes (clients and servers) and did not incorporate switches. This solution partially resolved the high load problem of network topologies, but the lack of resource isolation of switches has created performance problem in all the systems, and as a result, a network evaluation can produce inaccurate figures and results.

TABLE I. SUMMARY OF THE APPROACHES CHARACTERISTICS

| Features | Approaches | | | | |
|---|---|---|---|---|---|
|  | *[4]* | *[12]* | *[16]* | *[14]* | *Proposal* |
| Container Resource Isolation (all nodes) | ✗ | ✗ | ✓ | ✗ | ✓ |
| Container Resource Limitation (all nodes) | ✗ | ✗ | ✓ | ✗ | ✓ |
| Scalability | ✗ | ✓ | ✓ | ✗ | ✓ |
| Multi-Operating System (all nodes) | ✗ | ✗ | ✓ | ✗ | ✓ |
| SDN Protocol Support | ✓ | ✓ | ✗ | ✓ | ✓ |
| Legacy Protocol Support | ✗ | ✗ | ✓ | ✗ | ✓ |
| Real Network Behavior | ✓ | ✗ | ✓ | ✓ | ✓ |

Table I summarizes the characteristics of the solutions presented in this section. The characteristics are evaluated by features related to the solutions and compared with our proposal.

IV. vSDNEmul

A. vSDNEmul Overview Design

vSDNEmul is a container-based emulator that allows the construction of complex networks with realism and scalability. The base of the structure of vSDNEmul is the use of Docker containers. This solution facilitates building more complete containers compared with namespaces. This emu-lator also provides containers with memory, CPU or storage resources.

Our proposal provides ease in defining and modeling what resources each image will use, such as applications or tools. Those configurations are managed through files called DockerFiles. Thus, each node in the emulated network executes in an independent and isolated fashion, increasing the realism of the emulation and affording behavior similar to that in production infrastructures.

In addition, Docker enables containers with resource constraints. Each container is executed with memory, CPU, storage or network limitations. With this feature, nodes with resource requirements can be setup during the emulation, which allows the emulation of several types of behaviors in an experiment.

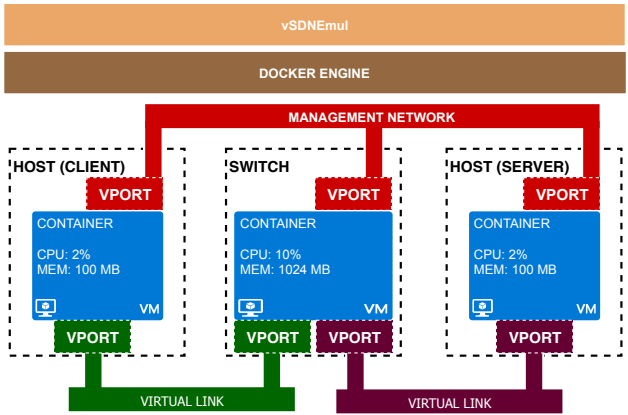

Figure 4. Overview of the vSDNEmul proposal.

Figure 4 depicts an overview of vSDNEmul. The overview shows the main components and the connections of the components in a simplified topology with 2 hosts and 1 switch created using vSDNEmul. Two API layers (Docker Engine and vSDNEmul) reside above the topology. These layers are used for implementing the proposal.

- *Container*: This component illustrates the nodes used by network topologies on the emulator. Each node is a Docker image with an explicit operating system and a set of tools that can describe a host, switch or router. The hosts can be one of two types: client or server. Clients generate several categories of traffic or consume network services, which are provided by servers. In turn, network devices (switches and routers) are implemented by means of switch and router software, such as OpenvSwitch. Therefore, the container component can also represent a network controller to which the switches connect during an experiment.
- *Virtual Port*: A virtual port is an element that emulates a network interface. Each node is composed of at least two virtual interfaces, where an interface is used to access the control and management of the nodes and one or more other ports are used for forwarding the data on emulated topology. These virtual interfaces are virtual Ethernet pair ports (veth devices) available on Linux O but could also be tunnel ports (TUN/TAP) resulting from GRE or VXLAN tunnels.
- *Virtual Link*: A virtual link is an emulator component responsible for creating interconnections among two or more nodes according to the topology a user has defined. There are two types of virtual links: point-to-point, which connects nodes, and point-to-multipoint, which creates a control and management bus among nodes. Theses virtual links can be either Ethernet virtual links or tunnels.

B. *vSDNEmul Architecture*

vSDNEmul is formed by a three-layer architecture that was designed to configure network device and computer machine characteristics. Based on the architecture of vSDNEmul, our work aims to introduce a new model for implementing a new network emulator. Figure 5 illustrates the architecture and respective layers: User, API and Infrastructure.

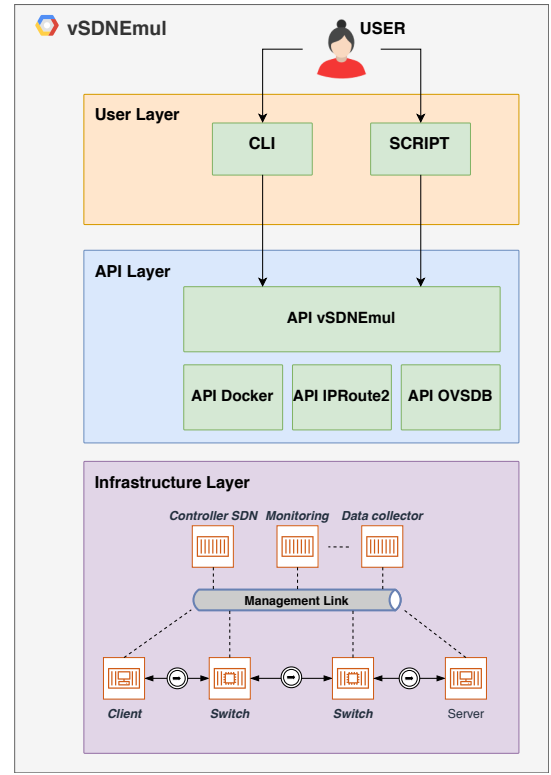

Figure 5. vSDNEmul architecture.

The User layer is the block that interacts with the user to build virtual networks. The user can set up her virtual network through CLI commands or Python scripts. The API layer comprises the libraries needed to perform resource allocation on the host computer, such as virtual links and machines. In addition, this layer contains the vSDNEmul API that abstracts low level elements on the emulator (e.g., a new node based on a Docker image or new links). This API has been implemented in the Python programming language with the purpose of easing maintenance and decreasing the learning curve of the emulator. The infrastructure layer consists of the logical elements of an emulation, such as clients, servers, controllers, interfaces, and links, that represent the topology of the experiment.

The goal of vSDNEmul is to provide emulation of the control and data planes as realistically as possible. To achieve this goal, the use of containers is essential. The main difference, when considering the operation of Mininet, is that all the nodes in vSDNEmul have isolated resources. As a result, the emulation fidelity is enhanced as each node operates independently, such as in physical infrastructures.

Another advantage of architecture of vSDNEmul is related to its scalability, which may be high or low according to the computing resources available. Thus, in low or medium-scale topologies, a local Docker installation can be used, while in large-scale topologies, a Docker cluster can be allocated using Kubernetes or Swarm clouds.

### C. Interacting with The Operation Modes

The vSDNEmul emulator enables two modes of operation that a user can employ to create and interact with the elements of a virtual network. In the following, we give brief descriptions of the operation modes:

- **CLI**: In CLI mode, all the information and configuration of the experiment are inserted by command line. Therefore, this mode is recommended only for small experiments. The user can execute only basic operations related to the network experiment, such as create, list, update, and delete elements. The user can apply these actions to links or nodes available on the data plane. Moreover, this mode allows for interaction with the command prompt of nodes to execute manual operations on applications or operating systems. In short, the CLI mode offers a few actions over the elements of the emulator. However, the user can also extend these actions through the vSDNEmul API;
- **Script**: In script mode, every topology customization is done by means of Python scripts using the vSDNEmul API. This mode is useful for complex experiments that require several configurations regarding hosts, switches, and links. This mode also allows the creation of events during the experiment that trigger several actions, such as disabling/enabling a port, pausing a node, executing a command and creating a node. This mode eases multiple repetitions of the same experiment.

### D. Customizing an Emulated Network

The vSDNEmul emulator exports a Python API that supports the production of custom topologies and experiments. This API includes an abstraction design and facilitates the creation of network components when setting up a network. This flexibility allows few lines of source code to both represent and describe a network, execute commands on various nodes, and collect results.

```python
dp = Dataplane()

# Adding SDN Switch
sw1 = dp.addNode(Whitebox(name="sw1"))

#Adding Two Hosts Clients
h1 = dp.addNode(Host(name="h1", ip="10.0.0.1" , mask="24"))
h2 = dp.addNode(Host(name="h2", ip="10.0.0.2", mask="24"))

#Creating Link Connection
# Link Between h1 to sw1
l1 = dp.addLink(LinkPair(name="l1",
                         node_source=sw1,
                         node_target=h1,
                         type=LinkType.HOST))
# Link Between h2 to sw2
l2 = dp.addLink(LinkPair(name="l2",
                         node_source=sw1,
                         node_target=h2,
                         type=LinkType.HOST))
# Creating a SDN Controller and setting to switch
ctl = dp.addNode(Onos(name="ctl1"))
mgnt = "tcp:{ip}:6653".format(ip=ctl.getIpController())
sw1.setController(target=mgnt, bridge="br_oper0")

#enabling cli
cli = Cli(dp)
cli.cmdloop()

#destroing all elements after the experiment.
dp.stop()
```

Figure 6. Basic example script of topology on vSDNEmul.

Figure 6 illustrates a script that sets up a small topology with one controller, two hosts, one switch and two links that connect the hosts to the switch. Essentially, the script includes a data plane environment to which the nodes are added (e.g., h1, h2, ctl and sw1). Subsequently, the link on the data plane is built to connect the hosts to the switch (e.g., h1 → sw1, h2 → sw2). Next, the script associates the SDN switch to the controller that dictates forwarding operations in the switch and permits the connection between h1 to h2. The last lines start the CLI and the experiment. Upon executing these operations, all elements and resources are deallocated.

### E. Node and Link Models

vSDNEmul uses containers to describe the nodes in emulated network topologies. The containers may comprise distinct elements during the emulation, such as host and network devices. With Docker, it is possible to customize the composition and execution of each node (e.g., the features of an operating system and the services executed can be customized). Our emulator implements this customization by reading the DockerFile that contains all the commands needed to build a custom image.

This structure allows vSDNEmul to work with node models of solutions updated or personalized by the users. In addition, the tests use the same software (i.e., the same version) as the production environment, making the scenarios of vSDNEmul more realistic than other solutions. Furthermore, Docker has a large community that shares images that can ease the deployment of new models and the reproducibility of other users' experiments. Currently, vSDNEmul provides a set of custom node models, described as follows:

- **Whitebox**: The whitebox node is an SDN switch based on Open vSwitch that enables management and control through the OpenFlow and OVSDB pro-tools. This node allows on-demand creation of virtual switches using the OVSDB protocol. The whitebox node is based on a novel SDN switch approach;
- **ONOS**: The ONOS node is an SDN controller based on an open-source controller, named ONOS (Open Network Operating System), that is considered the most robust SDN controller available, implementing new protocols for network controlling. ONOS enables control over several SDN protocols, such as OpenFlow, OVSDB, LISP, NETCONF and SNMP;
- **Host**: The host node is a basic container that implements client or server nodes on the emulated network. The client node is a container with tools to analyze network metrics (e.g., throughput, jitter, delay, bandwidth and round-rip). In turn, the server node is based on the real software of the network service application. Docker offers several network service containers for test and production scenarios. This node can support any host container available in the Docker library.

vSDNEmul implements link models to emulate wired connection among virtual interfaces. These interfaces forward data packets through hosts and switches. The available models can vary according to the host operating system. In the following, we describe two virtual link models available for users:

- **Veth**: Veth acts as a wired link that sets up a bridge between a container and either a physical network device or another container. This model can be used as a standalone network device. A Veth link emulates Ethernet ports or OSI layer 2 communications and is a simple way to interconnect Docker containers;
- **Tunnel**: Tunnels are wired links that can be used in two distinct scenarios: first, when a veth or similar link is not available in the operating system and an interconnection between two nodes is needed; second, when the node scale is the largest, more than one Docker host needs to be used for allocating nodes. The Tunnel model supported is based on GRE and VXLAN solutions.

In principle, our proposal employs wired links. However, the solution affords designing extensions for emulating wireless and optical links.

## V. RESULTS AND DISCUSSION

This section describes the methodology we used to evaluate vSDNEmul and compares the methodology and results obtained to those of Mininet. We specify two distinct evaluation methods. The first method analyzes the emulator performance regarding the number of nodes arranged in three types of topologies: tree, mesh and star. We investigate the following metrics: CPU use, amount of memory used, latency and throughput. The second method tests the fidelity of the emulators regarding throughput. Here, we set up a data center scenario with background traffic that stresses the queue ports. We aim to investigate any kind of unsatisfactory emulator behavior that could compromise the emulation and results.

### A. Evaluation Setup

Because Mininet is highly popular in the SDN academic community and has a high rate of citation in scientific papers, we selected Mininet for testing and collecting results. We used the latest version (2.3.0d5) of the emulator available on Mininet's GitHub repository.

The evaluation was performed using a KVM virtual machine with 2 CPU cores, 14 GB of RAM, 80 GB of storage capacity and Ubuntu OS server (18.04 is the oldest version that supports the minimum tools required to execute the operating system). The virtual machine was hosted by a Dell desktop with an Intel Core i7-4790 processor (4 cores), 24 GB of RAM, and 1 TB of storage capacity.

### B. Evaluation 1: Scalability

In this evaluation, we set up three types of topologies: tree, star and mesh, as depicted in Figure 7. Then, for each metric evaluated, we increased the number of nodes in the topology, from 9 to 513 nodes (513 is the greatest number of switches that is supported by both emulators without loss of data packets). During the experiments, we evaluated the following metrics: i) CPU use, ii) memory use, iii) latency, and iv) throughput.

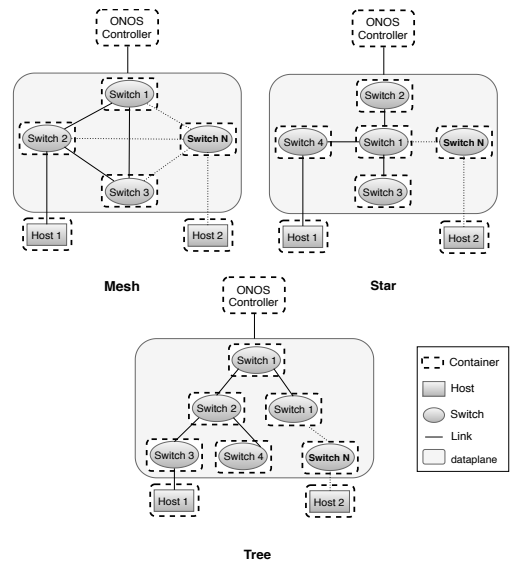

Figure 7. Types of topologies evaluated.

For metrics i and ii, we did not consider the use of SDN controllers, since the Mininet architecture does not provide SDN controllers for experiments. The result compilation was based on the topology components only (e.g., switches, links and hosts). Experiments were executed 15 times per test, and the final result was defined by calculating the average among all rates obtained.

TABLE II. CPU USE (%), WHERE THE CPU LIMIT IS 200%

| Switches | Mininet | | | vSDNEmul | | |
|---|---|---|---|---|---|---|
| | Star | Mesh | Tree | Star | Mesh | Tree |
| 9 | 1.06 | 1.14 | 1.11 | 2.70 | 2.72 | 2.71 |
| 17 | 1.78 | 2.35 | 2.13 | 5.56 | 6.50 | 6.31 |
| 33 | 2.82 | 3.88 | 3.95 | 15.65 | 26.21 | 17.05 |
| 65 | 3.93 | 6.63 | 4.65 | 41.32 | 35.85 | 32.42 |
| 129 | 18.25 | 26.67 | 20.10 | 36.47 | 42.82 | 38.23 |
| 257 | 37.01 | 41.06 | 38.45 | 83.63 | 92.26 | 86.16 |
| 513 | 51.27 | 55.5 | 52.75 | 169.35 | 182.13 | 173.22 |

TABLE III. MEMORY USE (MEGABYTES)

| Switches | Mininet | | | vSDNEmul | | |
|---|---|---|---|---|---|---|
| | Star | Mesh | Tree | Star | Mesh | Tree |
| 9 | 116.7 | 118.82 | 117.3 | 117.6 | 117.7 | 117.7 |
| 17 | 129.4 | 130.0 | 128.2 | 222.3 | 224.4 | 223.3 |
| 33 | 129.1 | 132.2 | 130.9 | 437.6 | 450.7 | 445.8 |
| 65 | 155.8 | 160.1 | 158.1 | 841.9 | 863.8 | 856.6 |
| 129 | 210.7 | 219.1 | 213.9 | 1725.2 | 1790.0 | 1754.0 |
| 257 | 317.1 | 323.5 | 318.7 | 3453.6 | 3651.7 | 3510.2 |
| 513 | 344.1 | 339.1 | 347.0 | 7121.8 | 7276.7 | 7237.4 |

TABLE IV. LATENCY FROM FIRST PING (MILISECOND)

| Switches | Mininet | | | vSDNEmul | | |
|---|---|---|---|---|---|---|
| | Star | Mesh | Tree | Star | Mesh | Tree |
| 9 | 18.7 | 20.8 | 24.9 | 16.7 | 16.0 | 17.6 |
| 17 | 29.2 | 29.9 | 30.8 | 24.5 | 25.5 | 26.0 |
| 33 | 63.6 | 74.2 | 56.6 | 58.1 | 59.1 | 55.2 |
| 65 | 77.2 | 89.8 | 73.9 | 88.6 | 89.6 | 70.3 |
| 129 | 143.1 | 181.1 | 144.0 | 154.3 | 159.3 | 129.0 |
| 257 | 344.2 | 380.3 | 337.5 | 359.1 | 364.1 | 294.4 |
| 513 | 650.2 | 726.0 | 672.5 | 628.3 | 635.3 | 609.9 |

Tables 2, 3 and 4 describe the results obtained from the evaluation of CPU use, memory use and latency, respectively, according to the topologies described above. As illustrated in Tables 2 and 3 Mininet performs better than vSDNEmul. The average memory used by vSDNEmul was on the order of 7 GB in all the topologies, while Mininet used only 347 MB. The CPU use for vSDNEmul was on the order of 147% of the free resources (i.e., 200%) available to the host. In turn, Mininet's CPU use was 55% for all the topologies. We stress that these results were expected due to the container solution implemented by the Docker framework. A Docker container has several management layers (network, kernel, process and mount). This additional management increased the memory and CPU use in vSDNEmul compared with the container solution used by Mininet (namespaces).

Table 4 presents latency results that represent the "average ping" found in each topology. Each of these values describes the time elapsed for forwarding a packet from a transmitter to a receiver (in the data plane) plus processing the packet in the network controller. We investigated this metric to study the network queue rate in both planes (data and control). We aim to compare the containers in the different approaches: the isolating approach (vSDNEmul) and the sharing approach (Mininet). As Table IV shows, vSDNEmul presents lower latency than Mininet, which confirms that the isolating approach of vSDNEmul is more efficient than the sharing approach for processing the network queue. This result is due to the sharing approach suffering from outside conflict from other processes in the operating system. In contrast, in the isolating approach, there is no outside interference. These aspects are further supported from analysis of the throughput results.

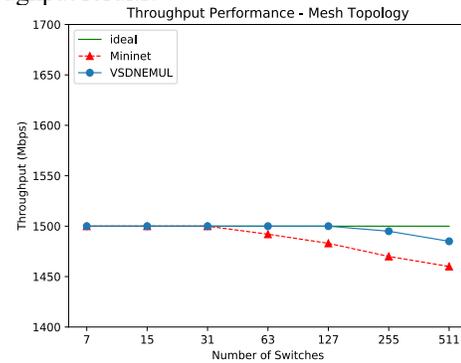

Figure 8. Throughput measured for mesh topology.

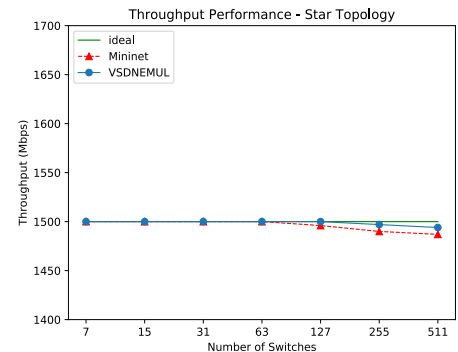

Figure 9. Throughput measured for star topology.

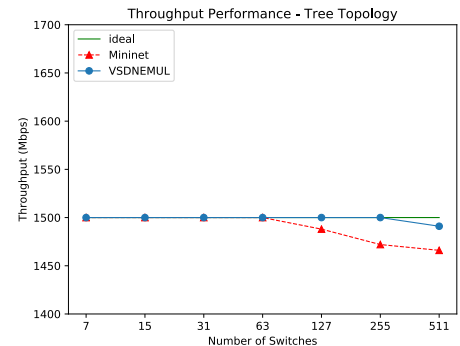

Figure 10. Throughput measured for tree topology.

Figures 8, 9 and 10 present the throughput reached when more nodes are added to the topology. The results indicate that increasing the number of nodes impacts the throughput for both emulators. Mininet, however, is not able to preserve the throughput close to the ideal proposed. This result is em-

phasized for topologies containing larger numbers of links, as shown in Figures 8 (mesh) and 10 (tree). Despite Mininet supporting many nodes, large numbers of links damage the throughput in the emulated topology and compromise the efficiency of the results. In contrast, vSDNEmul can support higher numbers of nodes while keeping the throughput closer to the ideal. When using the mesh, star and tree topologies, vSDNEmul supports 96, 64 and 192 more nodes than Mininet, respectively. Thus, these results confirm that vSDNEmul is able to support more nodes than Mininet without compromising accuracy.

## C. Evaluation 2: Realistic Behavior

In the second evaluation, we observed the emulator fidelity in terms of the amount of traffic that both proposals can emulate. We set up a scenario based on a tree topology, with 5 switches and 16 hosts, as illustrated in Figure 11. The purpose of this evaluation was to analyze the impact on the throughput during the emulation. To this end, we established a UDP iperf client (H1) and a server (H16), generating traffic at different rates of 1000, 1500, 2500 and 3000 Mbps. Then, we set up 7 client-server pairs in the other 14 remaining hosts (H2 to H15) to introduce background traffic at rates of 400, 600, 800, 1000 and 1200 Mbps. During the experiment, the accumulated traffic was within the maximum switching capacity measure, which was 6.6 Gbps.

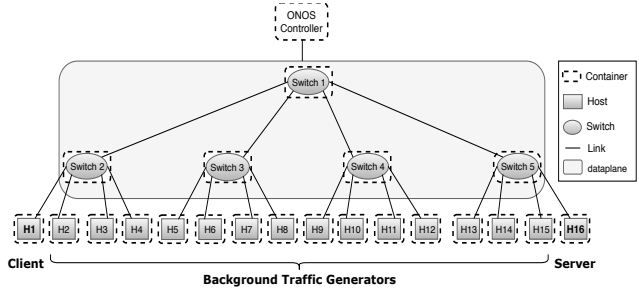

Figure 11. Topology used to measure the throughput performance between emulators.

Figures 12, 13, 14 and 15 present the results of experiments performed using evaluation method 2. We observed that in every scenario, Mininet could not control the foreground traffic at the desired rate (green line in all graphs). Additionally, for rates of 1000 and 1500 Mbps, this traffic moderately declined, which is illustrated in Figure 12. In turn, for higher rates, depicted in Figures 13, 14 and 15, the throughput declines abruptly. This result occurs due to Mininet's architecture, which forces a namespace to allocate all the network interfaces associated with the emulation. Therefore, for low traffic rates, there are no issues, but in scenarios with higher traffic rates, the concurrence in queue processing in the interfaces produces a bottleneck in the namespace processing that impacts the performance of all the network queues. As a result, the throughput is low in the experiment. Because this effect can result in false-positives during experiments with medium and high complexity, this issue largely limits the use of Mininet.

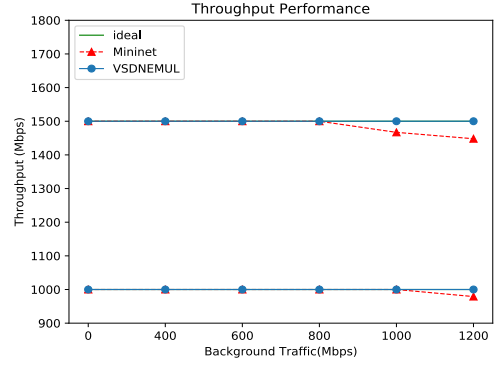

Figure 12. Throughput for rates of 1000 and 1500 Mbps.

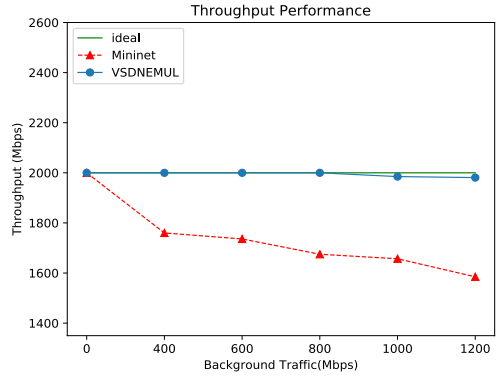

Figure 13. Throughput for rate of 2000 Mbps.

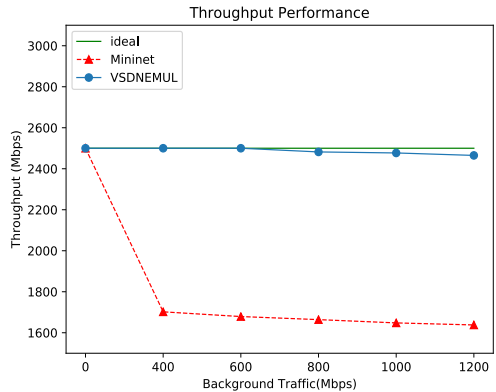

Figure 14. Throughput for rate of 2500 Mbps.

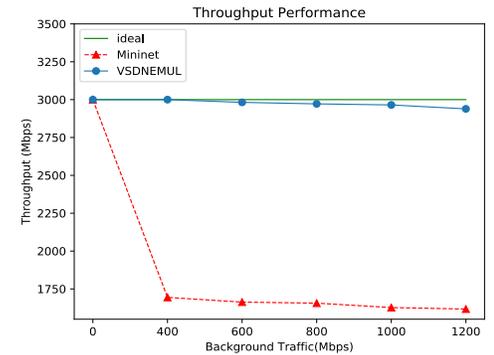

Figure 15. Throughput for rate of 3000 Mbps

Unlike Mininet, vSDNEmul manages the foreground traffic well, maintaining the throughput close to the desired rate. This outcome determines the level of reliability while using vSDNEmul for experiments. The results (Figures 12 – 15) also show that the isolation created by the Docker containers minimizes concurrence in queue processing, producing more reliable experiments and a network emulation with a good level of accuracy and realistic behavior.

## VI. CONCLUSION

This paper proposed vSDNEmul, an alternative solution SDN emulator that uses Docker containers to represent the elements in an emulated network. We introduced concepts related to SDN and container-based emulation. Additionally, we presented the design overview and architecture of our proposal and a comparative analysis between vSDNEmul and Mininet.

The results produced in the performance evaluation indicated that Mininet uses less computing resources than vSDNEmul to set up and maintain a network emulation, as mentioned in the related work. However, the latency and throughput are affected when the number of nodes increases in topologies with many links (e.g., mesh and tree). In addition, in a second evaluation, Mininet presented a crucial issue related to throughput that compromises the efficiency of the results obtained in experiments that use Mininet. The results showed that vSDNEmul provides more realistic and precise emulation results.

As future work, our proposal presents demand for more effectively controlling memory and CPU use through the application of more Docker hosts to allocate the containers. Thus, vSDNEmul could distribute the nodes and minimize the impacts on the memory and CPU during the emulation. Another issue related to vSDNEmul is scalability. A practical solution could be to integrate the emulator with cloud solutions such as Kubernetes, Swarm and Containerd and increase the number of nodes that could be allocated for an emulation. Additionally, in the context of cloud computing, our proposal could actuate intelligent solutions for discovering and allocating resources in cloud infrastructures and use these solutions in large-scale emulations.

ACKNOWLEDGMENT

This research was partially supported under the grant agreement no. 777067 (NECOS – *Novel Enablers for Cloud Slicing*), funded by the European Commission and the Brazilian Ministry of Science, Technology, Innovation, and Communication (MCTIC). The research also was financed in part by the Pró-Reitoria de Pesquisa e Pós-Graduação (Propesp/UFPA)

REFERENCES

[1] D. Kreutz, F. M. V Ramos, P. E. Veríssimo, C. E. Rothenberg, S. Azodolmolky, and S. Uhlig, "Software-Defined Networking : A Comprehensive Survey," *Proc. IEEE*, vol. 103, no. 1, pp. 14–76, 2015.

[2] B. A. A. Nunes, M. Mendonca, X. N. Nguyen, K. Obraczka, and T. Turletti, "A survey of software-defined networking: Past, present, and future of programmable networks," *IEEE Commun. Surv. Tutorials*, vol. 16, no. 3, pp. 1617–1634, 2014.

[3] F. Hu, Q. Hao, and K. Bao, "A survey on software-defined network and OpenFlow: From concept to implementation," *IEEE Commun. Surv. Tutorials*, vol. 16, no. 4, pp. 2181–2206, 2014.

[4] B. Lantz, B. Heller, and N. McKeown, "A network in a laptop," *Proc. Ninth ACM SIGCOMM Work. Hot Top. Networks - Hotnets '10*, pp. 1–6, 2010.

[5] P. Wette, M. Dräxler, and A. Schwabe, "MaxiNet: Distributed emulation of software-defined networks," in *2014 IFIP Networking Conference, IFIP Networking 2014*, 2014.

[6] R. R. Fontes, S. Afzal, S. H. B. Brito, M. A. S. Santos, and C. E. Rothenberg, "Mininet-WiFi: Emulating software-defined wireless networks," in *Proceedings of the 11th International Conference on Network and Service Management, CNSM 2015*, 2015, pp. 384–389.

[7] N. McKeown *et al.*, "OpenFlow," *ACM SIGCOMM Comput. Commun. Rev.*, vol. 38, no. 2, p. 69, Mar. 2008.

[8] E. Haleplidis, K. Pentikousis, S. Denazis, J. H. Salim, D. Meyer, and O. Koufopavlou, "Software-Defined Networking (SDN): Layers and Architecture Terminology," *RFC 7426*, 2015. [Online]. Available: https://rfc-editor.org/rfc/rfc7426.txt. [Accessed: 20-Sep-2001].

[9] A. Blenk, A. Basta, M. Reisslein, and W. Kellerer, "Survey on network virtualization hypervisors for software defined networking," *IEEE Commun. Surv. Tutorials*, vol. 18, no. 1, pp. 655–685, 2016.

[10] N. Handigol, B. Heller, V. Jeyakumar, B. Lantz, and N. McKeown, "Reproducible network experiments using container-based emulation," in *Proceedings of the 8th international conference on Emerging networking experiments and technologies - CoNEXT '12*, 2012, p. 253.

[11] B. Pfaff *et al.*, "The Design and Implementation of Open vSwitch," *Nsdi*, pp. 117–130, 2015.

[12] S. Y. Wang, C. L. Chou, and C. M. Yang, "EstiNet openflow network simulator and emulator," *IEEE Commun. Mag.*, vol. 51, no. 9, pp. 110–117, 2013.

[13] A. R. A. Kumar, S. V. Rao, and D. Goswami, "NS3 Simulator for a Study of Data Center Networks," in *2013 IEEE 12th International Symposium on Parallel and Distributed Computing*, 2013, pp. 224–231.

[14] M. Peuster, J. Kampmeyer, and H. Karl, "Containernet 2.0: A Rapid Prototyping Platform for Hybrid Service Function Chains," in *2018 4th IEEE Conference on Network Softwarization and Workshops (NetSoft)*, 2018, pp. 335–337.

[15] M. Hibler *et al.*, "Large-scale Virtualization in the Emulab Network Testbed," *Proc. 2008 USENIX Annu. Tech. Conf.*, pp. 113--128, 2008.

[16] A. Burtsev, P. Radhakrishnan, M. Hibler, and J. Lepreau, "Transparent checkpoints of closed distributed systems in emulab," in *Proceedings of the fourth ACM european conference on Computer systems - EuroSys '09*, 2009, p. 173